\documentclass[12pt,preprint]{aastex631}

\usepackage{amsmath}
\usepackage{amssymb}
\usepackage{graphicx}
\usepackage{natbib}

\usepackage{xcolor}
\newcommand{\rev}[1]{#1}

\begin{document}

\title{Triggering physical plasmoids in forming current sheets: conditions and diagnostics}

\author{H. Baty}
\affiliation{Observatoire Astronomique de Strasbourg, Universit\'e de Strasbourg, CNRS, UMR 7550, 11 rue de l'Universit\'e, F-67000 Strasbourg, France}
\correspondingauthor{H. Baty}
\email{hubert.baty@astro.unistra.fr}

\begin{abstract}
We investigate the conditions for triggering the plasmoid instability in a
dynamically forming current sheet in the resistive magnetohydrodynamic framework,
using a pseudo-spectral code applied to the Orszag-Tang vortex at Lundquist
number $S \sim 10^5$. Following Garc\'ia Morillo and Alexakis
[Garc\'ia Morillo and Alexakis, J. Fluid Mech. \textbf{1007}, R3 (2025)],
we use the power spectrum of the current density $E_J(k)$, complemented by the
vorticity spectrum $E_\omega(k)$, to assess the convergence of our simulations,
and show that this diagnostic remains valid even in the presence of physical
plasmoids, allowing us to unambiguously distinguish them from spurious ones. We
then show that physical plasmoids can be triggered in a well-resolved spectral
simulation when three conditions are simultaneously met: a perturbation applied
near the time of maximum current density, with amplitude above a critical
threshold $\varepsilon_c \sim 10^{-5}$ for our numerical scheme, and with
spectral content containing the unstable wavenumbers. These conditions are
confirmed using continuous noise injection, which yields similar results at
amplitudes one to two orders of magnitude lower. The resulting growth rates and
plasmoid numbers are in good agreement with the theory of Comisso et al.
[Comisso et al., ApJ \textbf{850}, 142 (2017)]. These results resolve the
apparent paradox raised by Garc\'ia Morillo and Alexakis (2025) and also clarify
the role of numerical noise in the triggering of the plasmoid instability.
\end{abstract}


\section{Introduction}

Magnetic reconnection is believed to be the underlying mechanism responsible for explosive energy release in many astrophysical and laboratory plasma environments. In the magnetohydrodynamic (MHD) framework, the classical Sweet-Parker model predicts reconnection rates scaling as $S^{-1/2}$ that are too slow to explain observed phenomena \citep{Priest2000}. For the Lundquist numbers expected in such plasmas, which can reach $S \sim 10^{10}$ or higher, this discrepancy amounts to many orders of magnitude. However, it has been realized that for Lundquist numbers exceeding a critical value of order $10^4$, reconnecting current sheets become unstable to the formation of plasmoid chains, leading to a fast reconnection regime with rates nearly independent of $S$ \citep{Loureiro2007, Bhattacharjee2009, Samtaney2009, HuangBhattacharjee2010}. A review of this plasmoid-dominated reconnection regime can be found in \citet{LoureiroPPCF2016}. Here $S = L_c v_{A,c} / \eta$, where $L_c$ and $v_{A,c}$ are a characteristic length scale and Alfv\'en velocity respectively, and $\eta$ is the resistivity. Different choices of $L_c$ are found in the literature. Here, following the seminal definition of \citet{Loureiro2007}, we take $L_c$ to be the current sheet length $L_{cs}$.

The linear theory of the plasmoid instability has been extensively studied. For a preformed static Sweet-Parker current sheet, having an aspect ratio scaling as $S^{1/2}$, the dominant wavenumber and growth rate scale as $S^{3/8}$ and $S^{1/4}$ respectively \citep{Loureiro2007}. More recently, \citet{Pucci2014} showed that current sheets with aspect ratio scaling as $S^{1/3}$ become linearly unstable on an Alfv\'enic time scale independent of $S$, providing an elegant criterion for the onset of fast reconnection. This result has been validated numerically, primarily using preformed current layers having an aspect ratio close to this expected critical value \citep[and references therein]{Tenerani2020}.

When current sheets form dynamically, either as a consequence of an ideal MHD instability or driven by an external velocity field, a different scenario is expected. For such forming current sheets, \citet{Comisso2016, Comisso2017} developed a general theory based on a principle of least time, showing that the scaling laws of the plasmoid instability are no longer simple power laws but depend on the Lundquist number. In particular, super-Alfv\'enic growth rates are obtained even for moderately large Lundquist numbers such as $S \sim 10^5$. These predictions have been confirmed in numerical simulations using the coalescence instability setup \citep{Huang2017}.

\citet{Alexakis2025} (hereafter GM\&A) recently performed a careful numerical study of the plasmoid regime (i.e. $S \gg 10^4$) in the Orszag-Tang vortex, a canonical MHD configuration in which a current sheet forms dynamically. Using a 2D pseudo-spectral code at very high resolution, they showed that numerical resolution plays a critical role: plasmoids observed at insufficient resolution are favored by the lack of resolution, which allows non-physical topological changes of field lines while failing to correctly capture their dynamics. A central result of their study is that the power spectrum of the current density $E_J(k)$ provides a robust and unambiguous diagnostic to distinguish well-resolved from under-resolved simulations. More surprisingly, they found that no physical plasmoids appear even at relatively large Lundquist numbers in their well-resolved simulations, raising the question of the conditions under which the plasmoid instability can actually be triggered.

A critical and somewhat overlooked aspect of these numerical studies concerns the role of numerical noise. Finite difference and finite element codes possess an intrinsic numerical noise that is orders of magnitude larger than that of spectral codes \citep{Ng2008}. In such codes, plasmoids can form spontaneously without any explicit perturbation, driven by this numerical noise. In spectral codes, on the contrary, numerical noise is extremely small, and plasmoids may simply not appear in a well-resolved simulation, not because the physical instability is absent, but because the noise level is insufficient to trigger it, or because there is not enough time for perturbations to grow to finite amplitude within the finite lifetime of the current sheet. This interpretation is indeed suggested by GM\&A themselves in their concluding remarks.

In this paper, we build upon the work of GM\&A and go further by systematically investigating the conditions under which physical plasmoids can be triggered in a well-resolved spectral simulation of the Orszag-Tang vortex. We restrict ourselves to a single value of the resistivity $\eta = 10^{-4}$, corresponding to $S \sim 10^5$, which is moderately large but sufficient to be well within the plasmoid regime, albeit lower than the values explored by GM\&A. Using a 2D pseudo-spectral code at resolutions ranging from $N = 1024$ to $N = 4096$, we first confirm that no plasmoids are observed in the absence of explicit perturbation, or when a perturbation is applied only at the initial time as in GM\&A. We then introduce controlled perturbations characterized by their amplitude $\varepsilon$, timing $t^*$, and spectral content (wavenumbers between $1$ and $k_{\rm sup}$), and show that all three parameters play a crucial role. 
We further validate these results by applying a continuous random perturbation,
which provides a more physically motivated noise model and confirms the robustness of the identified triggering conditions.
Our results are compared with the theoretical predictions of \citet{Comisso2017} and the numerical results of \citet{Huang2017}.

The paper is organized as follows. The numerical model and diagnostics are
described in Section~\ref{sec:model}. Section~\ref{sec:convergence} is devoted
to the convergence study and the role of power spectra as a diagnostic tool.
The physical triggering conditions are investigated in Section~\ref{sec:triggering}.
A comparison with theory and previous simulations is presented in
Section~\ref{sec:comparison}. Conclusions are drawn in Section~\ref{sec:conclusion}.

\section{Numerical model}
\label{sec:model}

\subsection{Equations, setup and numerical method}

We consider a reduced two-dimensional model of incompressible resistive MHD, in which all quantities are independent of the $z$-coordinate. The plasma density and magnetic permeability are both set to unity. The magnetic field and velocity field are expressed in terms of a magnetic vector potential $A$ and a stream function $\psi$ as \citep{Biskamp2000}
\begin{eqnarray}
\mathbf{B} = -\nabla \times (A\mathbf{e}_z) = \mathbf{e}_z \times \nabla A = (-\partial_y A, \, \partial_x A, \, 0), \\
\mathbf{v} = -\nabla \times (\psi\mathbf{e}_z) = \mathbf{e}_z \times \nabla \psi = (-\partial_y \psi, \, \partial_x \psi, \, 0),
\end{eqnarray}
where $\mathbf{e}_z$ is the unit vector perpendicular to the plane. Note that this convention differs from that of GM\&A, who use $\mathbf{B} = \nabla \times (a\mathbf{e}_z)$ and $\mathbf{v} = \nabla \times (\Psi\mathbf{e}_z)$, but the two formulations are physically equivalent. The current density $J$ and vorticity $\omega$ are the $z$-components of $\nabla \times \mathbf{B}$ and $\nabla \times \mathbf{v}$ respectively, giving $J = \nabla^2 A$ and $\omega = \nabla^2 \psi$. The governing equations, written with our convention, read

\begin{equation}
\frac{\partial \omega}{\partial t} + [\psi, \omega] = [A, J] + \nu \nabla^2 \omega,
\label{eq:omega}
\end{equation}

\begin{equation}
\frac{\partial A}{\partial t} + [\psi, A] = \eta \nabla^2 A,
\label{eq:A}
\end{equation}

where $[f, g] = \partial_x f \, \partial_y g - \partial_y f \, \partial_x g$ denotes the Poisson bracket.

The equations are solved in a double periodic domain of size $[0, 2\pi] \times [0, 2\pi]$ using a pseudo-spectral method on a uniform grid of $N \times N$ points, with a 4th-order Runge-Kutta scheme for time advancement and the 2/3 dealiasing rule. Simulations are performed at resolutions ranging from $N = 1024$ to $N = 4096$ with $\eta = \nu = 10^{-4}$, corresponding to a Lundquist number $S \sim 10^5$ (see Section~\ref{sec:convergence} for the precise estimate). Unless otherwise stated, the results presented in this paper are obtained at $N = 2048$. The numerical method and setup, including the initial conditions described below, are otherwise identical to those of GM\&A, except for the controlled perturbations introduced in Section~\ref{sec:triggering}.

The initial conditions are those of the Orszag-Tang vortex \citep{Orszag1979}:

\begin{equation}
A(t=0) = A_0 \left(-\cos x + \frac{1}{2}\cos 2y \right),
\end{equation}

\begin{equation}
\psi(t=0) = - \psi_0 \sin x \, \sin y,
\end{equation}

where $A_0 = 1$ and $\psi_0 = \sqrt{2}/2$, giving an initial magnetic energy density $\langle |\mathbf{B}|^2 \rangle / 2 = 1/2$ and kinetic energy density $\langle |\mathbf{v}|^2 \rangle / 2 = 1/8$.

\begin{figure}
\centering
\includegraphics[width=0.5\textwidth]{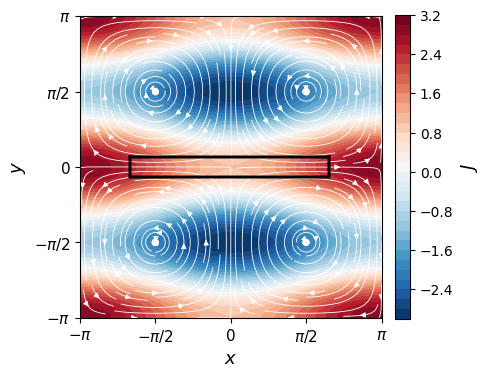}
\caption{Initial configuration of the Orszag-Tang vortex. Color map
shows the current density $J = \nabla^2 A$ and white streamlines
indicate the velocity field $\mathbf{v}$. The black rectangle
delimits the current sheet region along $y = 0$ that is the focus
of this study.}
\label{fig:OTini}
\end{figure}

The initial configuration of the Orszag-Tang vortex is illustrated in
Figure~\ref{fig:OTini}, which shows the current density $J$ together
with the velocity streamlines at $t = 0$. The velocity field exhibits
a pattern of converging stagnation-point flows directed toward the
$y = 0$ plane, which act to compress and thin the current layer
highlighted by the black rectangle. Simultaneously, the magnetic
field possesses an X-type null point structure at the origin, with
field lines being swept into the forming current sheet by the
overlying velocity field.

\subsection{Current sheet formation}

\begin{figure}
\centering
\includegraphics[width=0.99\textwidth]{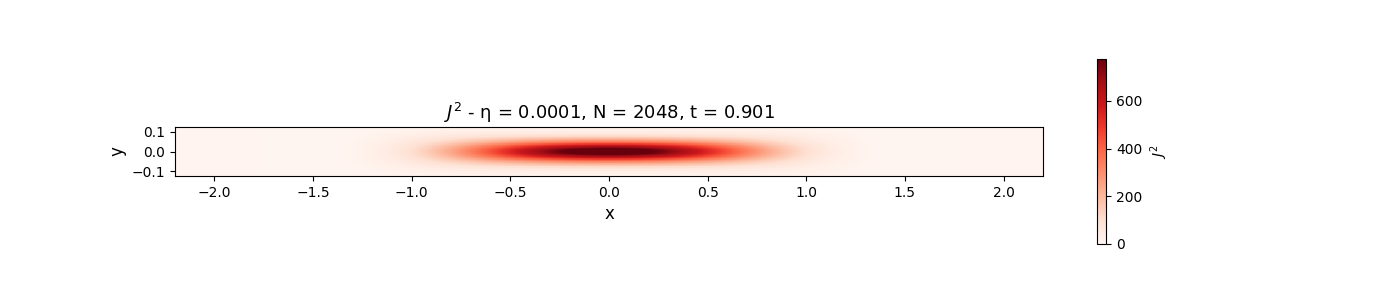}
\includegraphics[width=0.99\textwidth]{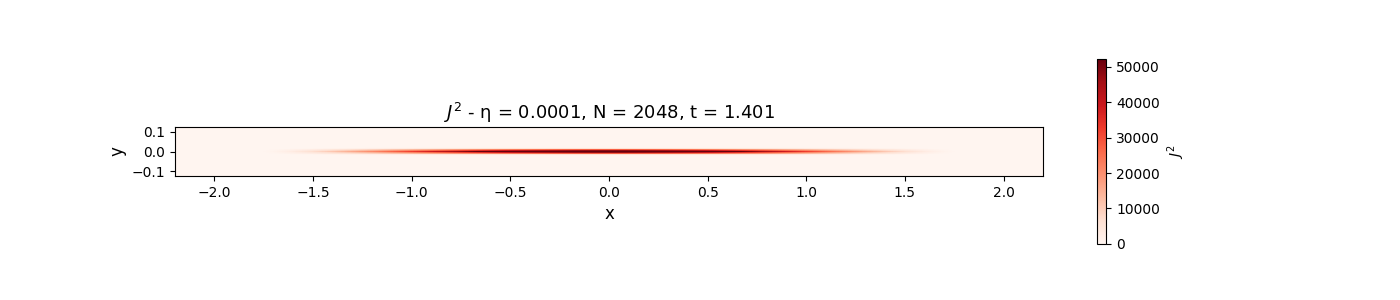}
\includegraphics[width=0.99\textwidth]{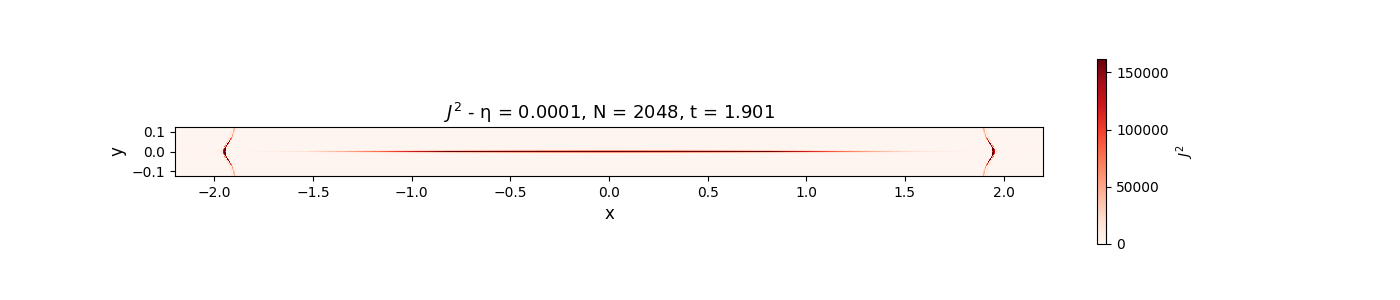}
\includegraphics[width=0.99\textwidth]{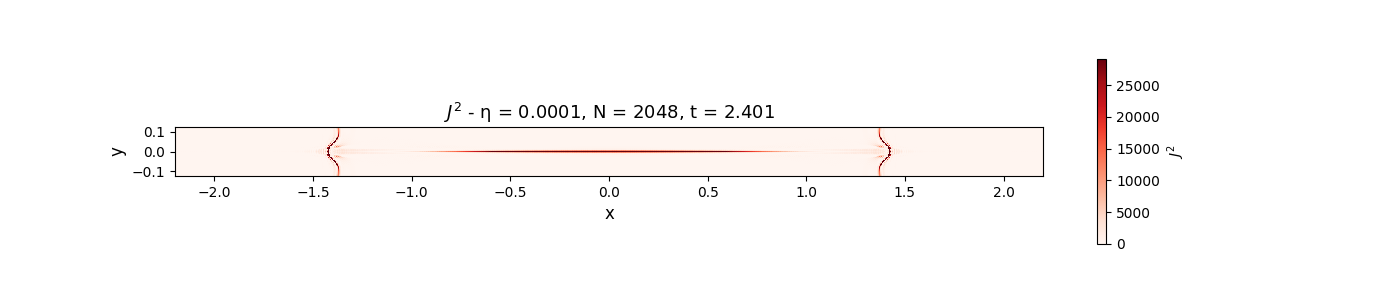}
\caption{Zoom onto the current sheet (region indicated by the black
rectangle in Figure~\ref{fig:OTini}) at $t = 0.9$, $1.4$, $1.9$,
and $2.4$ (from up to down). Color map shows the squared current density $J^2$. The sheet thins and
intensifies up to the peak at $t \simeq 1.9$, then relaxes.
$N = 2048$, $\eta = 10^{-4}$, and no perturbation ($\varepsilon = 0$). Times are expressed in Alfvén time units (see text for definition)}
\label{fig:sheet_evolution}
\end{figure}

The subsequent dynamical evolution leads to the progressive thinning
and intensification of the current sheet, as illustrated in
Figure~\ref{fig:sheet_evolution}, which shows a zoom onto the current
layer at four representative times: $t = 0.9$, $1.4$, $1.9$, and
$2.5$. During the first phase ($t \lesssim 1.9$), the sheet thins
continuously as the upstream flow compresses the layer, causing the
peak current density to grow. The current density reaches its maximum at $t \simeq 1.9$,
in agreement with GM\&A, marking the most favorable
instant for the onset of the plasmoid instability. This corresponds to a timescale $t_{\rm peak} \simeq 1.9 \sim \tau_A$ (see below), where $\tau_A = L_{cs}/v_A$ is the Alfv\'en time. The current sheet length $L_{cs}$ also reaches its maximum value
$L_{cs} \simeq 3$ at the same time. Beyond this time,
the driving velocity field weakens and the sheet begins to relax, broadening and diffusing thereafter.

\begin{figure}
\centering
\includegraphics[width=\textwidth]{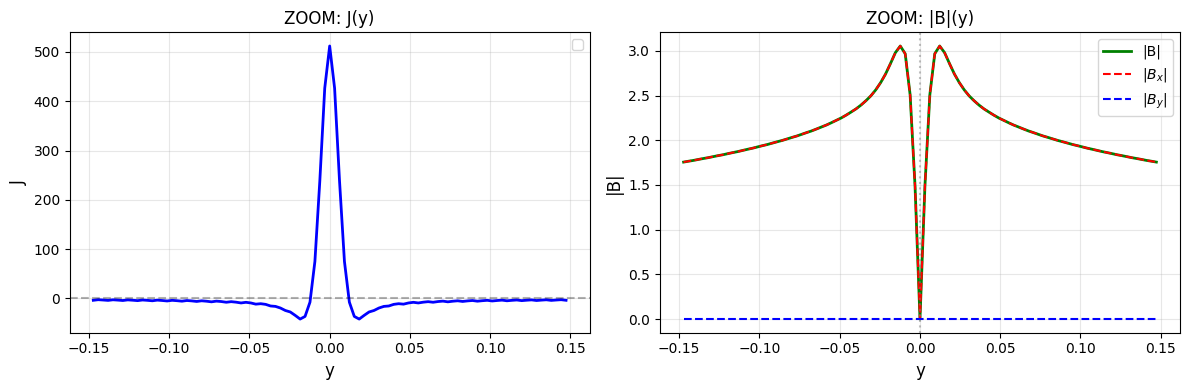}
\caption{Transverse profiles of the current density $J$ (left panel) and
magnetic field components (right panel) at $x = 0$ and $t = 1.9$.
The half-thickness $\delta$ and upstream Alfv\'en velocity $v_A$
can be extracted from these profiles, and used to estimate relevant parameters as the Lundquist
number $S$ (see text).
$N = 2048$, $\eta = 10^{-4}$, and $\varepsilon = 0$.}
\label{fig:cuts}
\end{figure}

Transverse profiles of $J$ and the magnetic field components across the sheet at $x = 0$
are shown in Figure~\ref{fig:cuts} at $t = 1.9$, i.e.\ at the
current density peak, providing a quantitative characterization of
the sheet geometry at its most critical stage. From these profiles,
the half-thickness $\delta \simeq 6 \times 10^{-3}$ (measured as half-width at half-maximum) and the upstream Alfv\'en velocity
$v_A = B_{up}$ (where $B_{up} \simeq 3$ is the evaluated reconnecting field) can be extracted, allowing in particular an estimate of the
Lundquist number $S = L_{cs}  v_A / \eta \simeq 10^5$. The resulting aspect ratio $L_{cs} / (2\delta) \simeq 250$, consistent with the Sweet-Parker scaling corrected for the magnetic Prandtl number $S^{1/2}/(1+P_m)^{1/4} \simeq 266$ (with $P_m = \nu/\eta = 1$) \citep{ComissoGrasso2016}, confirming that the current sheet geometry is consistent with Sweet-Parker theory at its peak. In order to estimate the triggering conditions for the plasmoid instability, we note that the Alfv\'en time $\tau_A = L_{cs}  / v_A \simeq 1$, so that the Alfv\'en time scale is close to unity in our units.

In order to investigate the triggering conditions for the plasmoid instability,
controlled perturbations are introduced at a chosen time $t^*$ by adding to the
vector potential a random noise of amplitude $\varepsilon$ filtered to
wavenumbers $k \leq k_{\rm sup}$:
\begin{equation}
A \rightarrow A + \varepsilon \cdot \xi(\mathbf{x}, k_{\rm sup}),
\end{equation}
\rev{where $\xi$ is a random field generated in physical space, normalized such that
$\max |\xi| = 1$, and filtered in Fourier space to retain only modes with
$k \leq k_{\rm sup}$. The amplitude $\varepsilon$ is relatively small compared
to the amplitude of $A$. In contrast to GM\&A, who apply a fixed perturbation
at $t = 0$, the amplitude $\varepsilon$ is here treated as a free parameter. We
have verified on two representative cases (single injection at $t^*$ and
continuous perturbation) that generating the perturbation directly in Fourier
space with random phases yields identical results. The three parameters
$\varepsilon$, $t^*$, and $k_{\rm sup}$ fully characterise the perturbation and
their respective roles are investigated systematically in the following sections.}

\section{Convergence and spurious plasmoids}
\label{sec:convergence}

The sensitivity of the plasmoid instability to numerical resolution in the Orszag-Tang
configuration has been recently investigated by GM\&A, who demonstrated
that well-resolved pseudo-spectral simulations display no plasmoids even at Lundquist
numbers as large as $S_{\rm A} \sim 5 \times 10^5$ with their definition
$S_{\rm A} \equiv B_{\rm max}/(\eta k_1)$ (where $k_1 = 1$ is the smallest non-zero
wavenumber). With our definition $S = L_{cs}\,v_A/\eta$, taking $L_{cs} \simeq 3$ and $v_A \simeq 3$, this corresponds to $S \sim 10^6$, significantly larger than the value $S \sim 10^5$ achieved in the present study at $\eta = 10^{-4}$.
Under-resolved runs in GM\&A produce spurious plasmoids. In this
section, we confirm these findings using our own pseudo-spectral code at $\eta = 10^{-4}$,
and we establish the power spectrum of the current density $E_J(k)$ as a reliable
convergence diagnostic. We further show that the vorticity spectrum $E_\omega(k)$ provides an additional and independent confirmation.

\subsection{Effect of numerical resolution on current sheet structure}
\label{sec:resolution}

\begin{figure}
\centering
\includegraphics[width=0.97\textwidth]{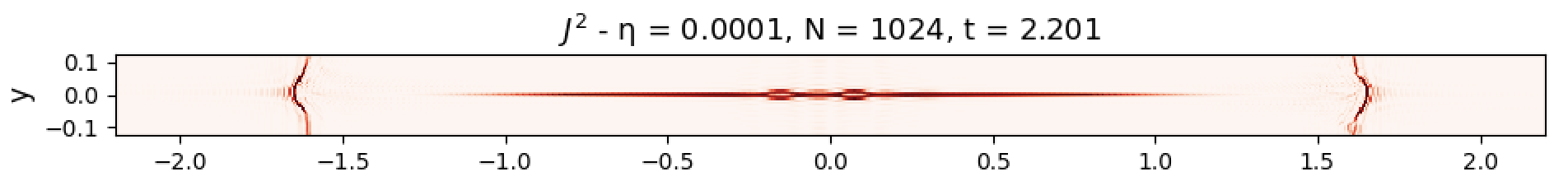}
\includegraphics[width=0.97\textwidth]{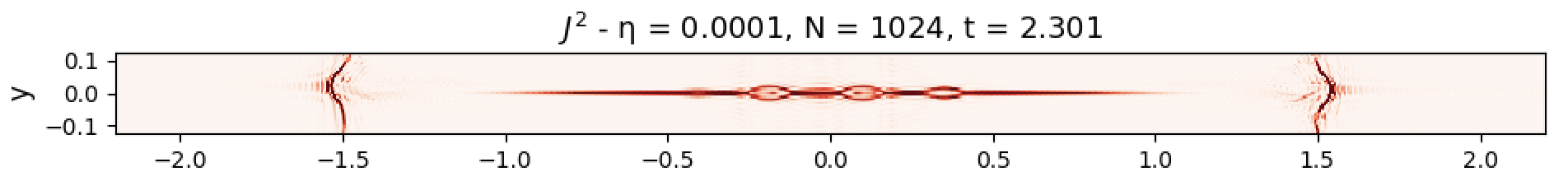}
\includegraphics[width=0.97\textwidth]{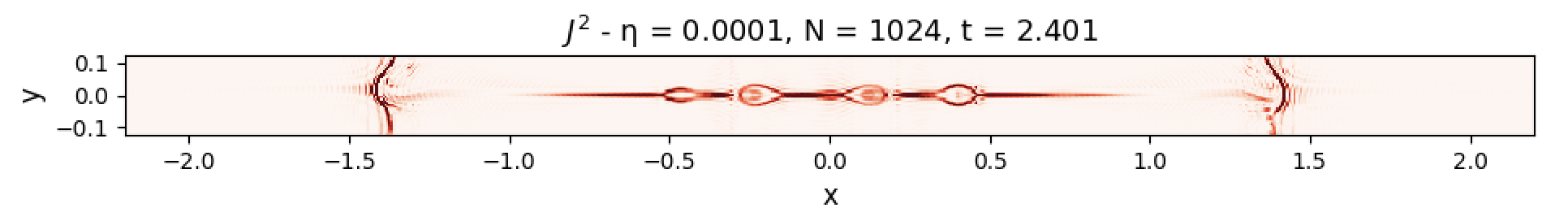}
\caption{Squared current density $J^2$ in the current sheet region at $t = 2.2$,
$2.3$, and $2.4$ (top to bottom) for $N = 1024$, $\eta = 10^{-4}$, and
$\varepsilon = 0$. Spurious plasmoids develop and grow during the relaxation phase,
in stark contrast with the smooth, plasmoid-free sheet observed at the same times
for the converged case $N = 2048$ (Figure~\ref{fig:sheet_evolution}). Times are expressed in Alfvén time units (see text for definition)}
\label{fig:sheets_resolution}
\end{figure}

Figure~\ref{fig:sheets_resolution} shows the squared current density $J^2$ in the
current sheet region at three successive times during the relaxation phase
($t = 2.2$, $2.3$, and $2.4$) for $N = 1024$, $\eta = 10^{-4}$, and $\varepsilon = 0$.
The reference converged case at $N = 2048$ has already been shown in
Figure~\ref{fig:sheet_evolution}, where the current sheet remains smooth and
featureless throughout. In stark contrast, at $N = 1024$ the sheet develops
well-defined plasmoid structures during the relaxation phase: magnetic islands
form and grow as the current layer broadens after the peak. Crucially, these plasmoids appear only because the simulation is under-resolved, which allows non-physical topological changes of field lines while failing to correctly capture their dynamics (GM\&A). As noted by GM\&A, this manifests itself with the formation of larger plasmoids the larger the under-resolving is. These results unambiguously establish that the plasmoids observed at low resolution are favored and poorly described by insufficient resolution.

\subsection{Power spectrum as a convergence diagnostic}
\label{sec:spectrum_convergence}

A quantitative and unambiguous convergence diagnostic is provided by the power
spectrum of the current density, $E_J(k) = k^2 E_B(k)$, where $E_B(k)$ is the
magnetic energy spectrum. \rev{Here $k = \sqrt{k_x^2 + k_y^2}$ is the radial
wavenumber}.
Following GM\&A, we consider a simulation
to be well-resolved if the peak of $E_J(k)$ is at least ten times larger than its
value at the maximum retained wavenumber $k_{\max} = N/3$, i.e.\ $\max_k\{E_J(k)\}
\geq 10\, E_J(k_{\max})$. This criterion ensures that most of the Ohmic dissipation
is correctly captured and that the high-$k$ modes do not carry a significant fraction
of the total current energy. When this condition is not met, the energy at the largest
wavenumbers is comparable to the spectral peak, indicating that the dissipative cascade
is not fully resolved and that the truncated Fourier modes can seed spurious instabilities.

\begin{figure}
\centering
\includegraphics[width=0.48\textwidth]{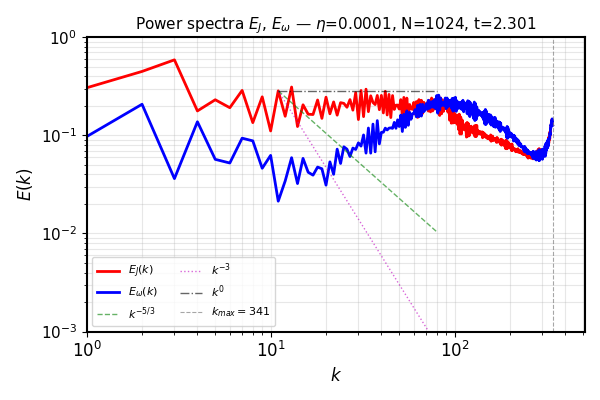}
\includegraphics[width=0.48\textwidth]{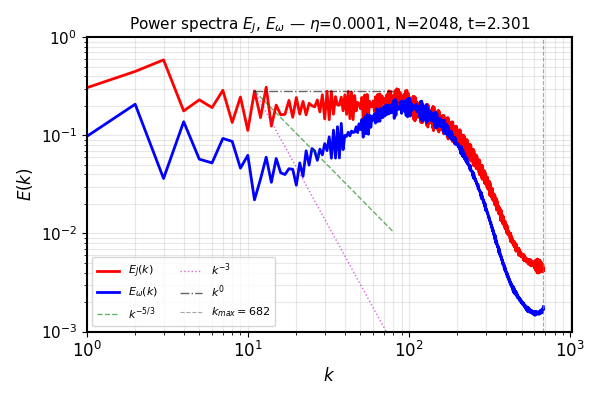}
\caption{Power spectra of the current density $E_J(k)$ (blue) and vorticity
$E_\omega(k)$ (red) at $t = 2.3$, for $N = 1024$ (left) and $N = 2048$ (right),
at $\eta = 10^{-4}$ and $\varepsilon = 0$. At $N = 1024$, both spectra fail the
convergence criterion $\max_k\{E_J(k)\} \geq 10\,E_J(k_{\max})$, indicating an
incomplete dissipative cascade in both fields (spurious plasmoids are present, see figure 4). At $N = 2048$, both spectra are
well-resolved, with energy decreasing to negligible values before $k_{\max}$ (plasmoids are absent, see figure 2).}
\label{fig:spectra_resolution}
\end{figure}

Figure~\ref{fig:spectra_resolution} shows both $E_J(k)$ and the vorticity spectrum
$E_\omega(k)$ at $t = 2.3$ for $N = 1024$ (left panel) and $N = 2048$ (right panel),
all at $\eta = 10^{-4}$ and $\varepsilon = 0$. At $N = 1024$, both spectra fail the
convergence criterion: the energy near $k_{\max}$ remains a significant fraction of
the spectral peak, reflecting an incomplete dissipative cascade for both the magnetic
and velocity fields. The high-$k$ modes effectively act as a broadband numerical noise
source of non-negligible amplitude, which seeds the spurious plasmoids visible in
Figure~\ref{fig:sheets_resolution}. The vorticity spectrum $E_\omega(k)$ provides an
independent and complementary confirmation of this diagnostic: the same under-resolution
signature is present in both fields simultaneously, reinforcing the conclusion that the
simulation is not converged. At $N = 2048$, both $E_J(k)$ and $E_\omega(k)$ satisfy
the convergence criterion and decrease by about two orders of magnitude before $k_{\max}$,
confirming that the simulation is well-resolved.

We note that GM\&A introduced a small random perturbation
($\sim 0.25\%$ of the total energy, restricted to modes $|k| < 16$) at $t = 0$
in their simulations in order to break the symmetry of the Orszag-Tang initial
conditions. We have verified that the inclusion or omission of such an initial
perturbation has no impact on the convergence behaviour described above: the
resolution threshold for the appearance of spurious plasmoids and the shape of
the power spectra are unaffected at $N = 2048$.

We also note that GM\&A explored Lundquist numbers significantly
higher than those studied here, reaching $S \sim 5 \times 10^5$ at resolutions
up to $N = 32768$. Such extreme resolutions are beyond the scope of the present
study, which is limited to $N \leq 4096$ with $\eta = 10^{-4}$.

In the following section, we go beyond the convergence study of GM\&A
and investigate the conditions under which physical plasmoids can be triggered in
a well-resolved simulation ($N = 2048$) by applying controlled perturbations at
chosen times $t^*$, with varying amplitude $\varepsilon$ and spectral content
$k_{\rm sup}$.

\section{Physical triggering conditions}
\label{sec:triggering}

Having established that a well-resolved pseudo-spectral simulation produces no
plasmoids in the absence of explicit perturbation, we now investigate the conditions
under which physical plasmoids can be triggered. All simulations in this section
are performed at $N = 2048$ and $\eta = 10^{-4}$, ensuring that the convergence
criterion is satisfied throughout. Selected cases are also repeated at $N = 4096$
to confirm the robustness of the results. Controlled perturbations of the form described
in Section~\ref{sec:model} are applied at time $t^*$, and their effect is
characterized through three parameters: the perturbation timing $t^*$, amplitude
$\varepsilon$, and spectral content $k_{\rm sup}$.

\subsection{Effect of perturbation timing $t^*$}

We first examine the role of the perturbation timing $t^*$, keeping the amplitude
$\varepsilon = 10^{-3}$ and spectral content $k_{\rm sup} = 32$ fixed. As
established in Section~2, the current sheet thins and intensifies up to
$t \simeq 1.9$, after which it relaxes. One therefore expects the plasmoid
instability to be most easily triggered when the perturbation is applied close
to the peak, where the sheet is thinnest and the local Lundquist number is
largest.

Table~\ref{tab:timing} summarizes the number of plasmoids $N_p$ and the measured
growth rate $\gamma$ for perturbations applied at $t^* = 1.5$, 1.7, and 1.9, as
well as the reference case $t^* = 0$ with $\varepsilon = 10^{-3}$, which
reproduces the setup of GM\&A. In this reference case, no plasmoids are observed,
confirming that an early perturbation is insufficient to trigger the instability,
in agreement with GM\&A. When a perturbation is applied later, plasmoids develop
in all three cases, with both $N_p$ and $\gamma$ increasing monotonically with
$t^*$. At $t^* = 1.5$, the sheet is still relatively thick and 9 plasmoids are
observed with $\gamma\tau_A \simeq 7.5$. At $t^* = 1.7$ and $t^* = 1.9$, the
sheet is thinner and the instability is stronger, yielding 10 and 11 plasmoids
with $\gamma\tau_A \simeq 9$ in both cases. The near-identical values at
$t^* = 1.7$ and $t^* = 1.9$ suggest that the instability is close to saturation
near the peak of the current sheet. \rev{The procedures used to measure $\gamma$ and
$N_p$ are detailed in Appendix~\ref{app:diagnostics}.}

\begin{table*}
\centering
\caption{Effect of perturbation timing $t^*$ on the number of plasmoids $N_p$
and measured growth rate $\gamma$, for $\varepsilon = 10^{-3}$, $k_{\rm sup} = 32$,
$N = 2048$, $\eta = 10^{-4}$. The case $t^* = 0$ reproduces the setup of
GM\&A and is shown for reference. Uncertainties on $N_p$ are $\pm 10\%$.
\label{tab:timing}}
\begin{tabular}{ccc}
\hline\hline
$t^*$ & $N_p$ & $\gamma \tau_A$ \\
\hline
0 (as in GM\&A) &  0 & --- \\
1.5                               &  9 & 7.5 \\
1.7                               & 10 & 9.0 \\
1.9                               & 11 & 9.0 \\
\hline
\end{tabular}
\end{table*}

The development of plasmoids for the case $t^* = 1.5$, $k_{\rm sup} = 32$ is
illustrated in Figure~\ref{fig:plasmo_kmax32}, which shows $J^2$ in the current
sheet region at $t = 1.8$ (plasmoids disrupting the sheet) and $t = 2.0$
(well-developed plasmoids), together with the power spectra $E_J(k)$ and
$E_\omega(k)$ at $t = 1.9$. At $t = 1.8$, small-scale magnetic islands are
already visible, disrupting the current sheet structure. By $t = 2.0$, a chain
of well-separated plasmoids has formed. Crucially, the power spectra at $t = 1.9$
satisfy the convergence criterion, with the ratio
$\max_k\{E_J(k)\}/E_J(k_{\max})$ slightly above 10 for the current density and
just at 10 for the vorticity (marginally satisfied), confirming that these
plasmoids are physical and not spurious. We note that the exact choice of time at
which the spectra are evaluated has little influence on this conclusion; $t = 1.9$
is chosen as a time at which the magnetic islands are sufficiently well developed.
The robustness of this conclusion will be further confirmed at higher resolution
$N = 4096$ in Section~\ref{sec:spectral}.

\subsection{Effect of perturbation amplitude $\varepsilon$}
\label{sec:amplitude}

We next investigate the role of the perturbation amplitude $\varepsilon$, fixing
$t^* = 1.5$ and $k_{\rm sup} = 32$. The results are summarized in
Table~\ref{tab:amplitude}. As $\varepsilon$ is decreased from $10^{-3}$ to
$10^{-6}$, both the number of plasmoids and the measured growth rate decrease
monotonically. At $\varepsilon = 10^{-3}$, a well-developed chain of $9$ plasmoids
is obtained with $\gamma \tau_A \simeq 7.5$. Reducing the amplitude to $\varepsilon =
10^{-4}$ yields a weaker response: $5$ plasmoids with a significantly reduced
growth rate $\gamma \tau_A \simeq 4$. At $\varepsilon = 10^{-5}$, only $2$ barely
visible, late-forming islands appear, with no measurable exponential growth
($\gamma \tau_A \simeq 0$), indicating that the perturbation is marginally sufficient
to trigger a nonlinear response. Finally, at $\varepsilon = 10^{-6}$, no magnetic
islands develop at all, confirming the existence of a critical threshold
$\varepsilon_c \sim 10^{-5}$ for our numerical scheme.
Indeed, one must note that the precise value of $\epsilon_c$ may depend on the specifics of the numerical implementation, which determine the effective noise floor of the spectral code.

This threshold is physically meaningful: it represents the minimum noise level
required for the perturbation to amplify from its initial amplitude to the
nonlinear regime within the finite lifetime of the current sheet. Below
$\varepsilon_c$, even though the linear instability criterion may be satisfied,
the perturbation simply does not have enough time to reach finite amplitude before
the sheet relaxes. This finite-time constraint is a key feature of forming current
sheets that distinguishes them from the idealized static configurations considered
in classical plasmoid instability theory.

\begin{table*}
\centering
\caption{Effect of perturbation amplitude $\varepsilon$ on the number of
plasmoids $N_p$ and measured growth rate $\gamma$, for $t^* = 1.5$,
$k_{\rm sup} = 32$, $N = 2048$, $\eta = 10^{-4}$. Uncertainties on $N_p$ are $\pm 10\%$.
\label{tab:amplitude}}
\begin{tabular}{cccc}
\hline\hline
$\varepsilon$ & $N_p$ & $\gamma \tau_A$ & Comments \\
\hline
$10^{-3}$ & 9 & 7.5          & well-developed chain \\
$10^{-4}$ & 5 & 4.0          & reduced activity \\
$10^{-5}$ & 2 & $\simeq 0$  & marginal, late-forming islands \\
$10^{-6}$ & 0 & ---          & no islands \\
\hline
\end{tabular}
\end{table*}

\subsection{Effect of spectral content $k_{\rm sup}$}
\label{sec:spectral}

We finally examine the role of the spectral content of the perturbation by varying
$k_{\rm sup}$ from $16$ to $256$, keeping $\varepsilon = 10^{-3}$ fixed. Results are
presented for two perturbation times, $t^* = 1.5$ and $t^* = 1.9$, in
Table~\ref{tab:kmax}.

For $t^* = 1.5$, both $N_p$ and $\gamma \tau_A$ increase monotonically with $k_{\rm sup}$
up to $k_{\rm sup} = 128$, beyond which saturation is reached, with maximum $N_p$ value of order $22$
and maximum $\gamma \tau_A$ of order $10$. At the lowest spectral content ($k_{\rm sup} = 16$), only $3$ plasmoids develop with
$\gamma \tau_A \simeq 5$, while at $k_{\rm sup} = 64$ already $18$ plasmoids are obtained
with $\gamma \tau_A \simeq 10$.
For $t^* = 1.9$, a behavior similar to $t^* = 1.5$ case is observed. However, the saturated values for $k_{\rm sup} = 128$ are 
slightly higher as now $N_p \simeq28$ and $\gamma \tau_A \simeq12$. The saturation observed for  $k_{\rm sup} = 128$ is physically significant and will be discussed in the context of the theoretical predictions in Section~\ref{sec:comparison}.

\begin{table*}
\centering
\caption{Effect of spectral content $k_{\rm sup}$ on the number of plasmoids $N_p$
and measured growth rate $\gamma$, for $t^* = 1.5$ and $t^* = 1.9$,
with $\varepsilon = 10^{-3}$, $N = 2048$, $\eta = 10^{-4}$.
Uncertainties on $N_p$ are $\pm 10\%$.
\label{tab:kmax}}
\begin{tabular}{cc@{\hspace{1.5em}}cc@{\hspace{1.5em}}cc}
\hline\hline
 & \multicolumn{2}{c}{$t^* = 1.5$} & \multicolumn{2}{c}{$t^* = 1.9$} \\
\cline{2-3}\cline{4-5}
$k_{\rm sup}$ & $N_p$ & $\gamma \tau_A$ & $N_p$ & $\gamma \tau_A$ \\
\hline
 16 &  3 &  5.0 &  3 &  6.5 \\
 32 &  9 &  7.5 & 11 &  9.0 \\
 64 & 18 & 10.0 & 21 & 11.0 \\
128 & 22 & 10.0 & 28 & 12.0 \\
256 & 22 & 10.0 & 28 & 12.0 \\
\hline
\end{tabular}
\end{table*}

\begin{figure}
\centering
\includegraphics[width=0.99\textwidth]{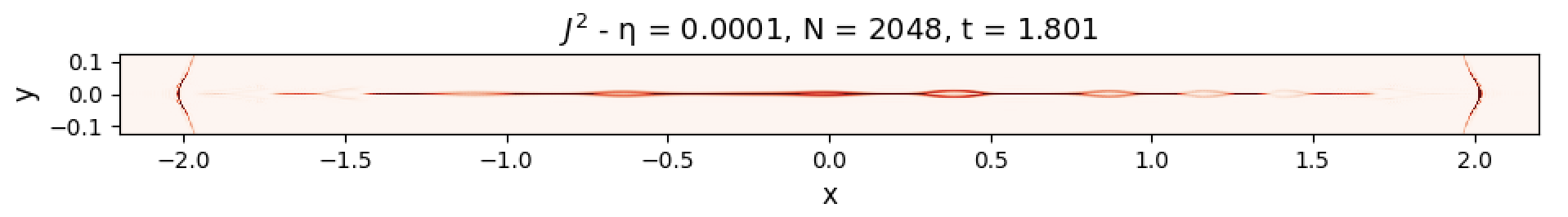}
\includegraphics[width=0.99\textwidth]{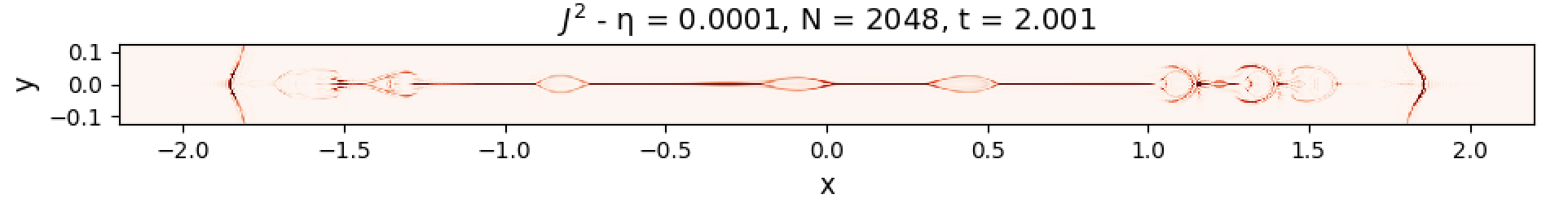}
\includegraphics[width=0.5\textwidth]{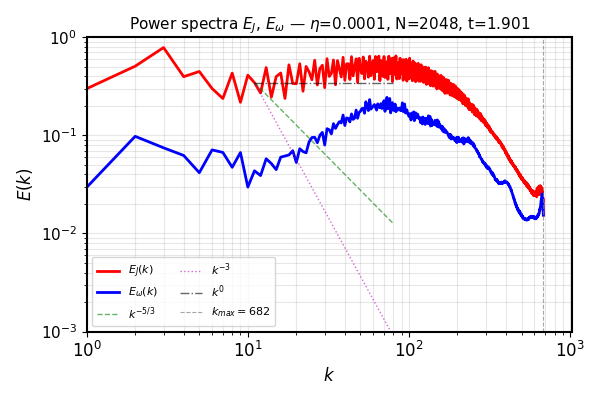}
\caption{Physical plasmoids triggered by a perturbation at $t^* = 1.5$ with
$\varepsilon = 10^{-3}$ and $k_{\rm sup} = 32$. Top and middle panels: squared
current density $J^2$ at $t = 1.8$ (onset) and $t = 2.0$ (well-developed
plasmoids). Bottom panel: power spectrum $E_J(k)$ at $t = 1.9$, satisfying
the convergence criterion and confirming the physical nature of the plasmoids.
$N = 2048$. Times are expressed in Alfvén time units (see text for definition).}
\label{fig:plasmo_kmax32}
\end{figure}

\begin{figure}
\centering
\includegraphics[width=0.99\textwidth]{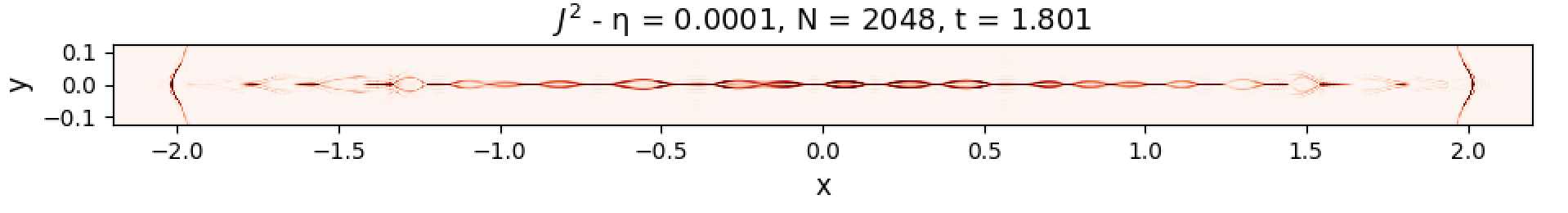}
\includegraphics[width=0.99\textwidth]{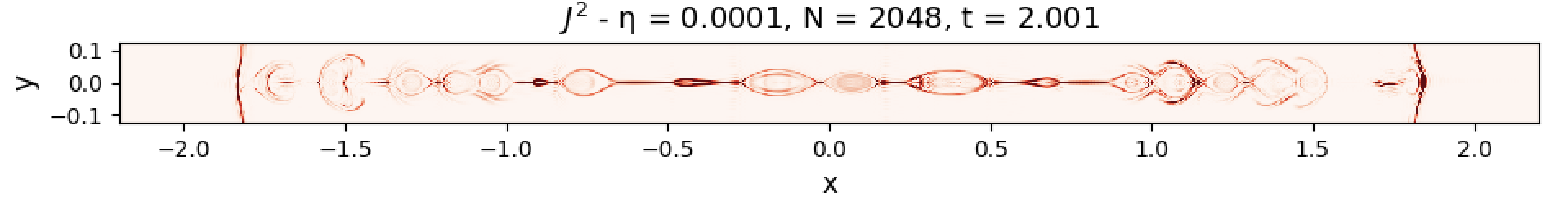}
\includegraphics[width=0.5\textwidth]{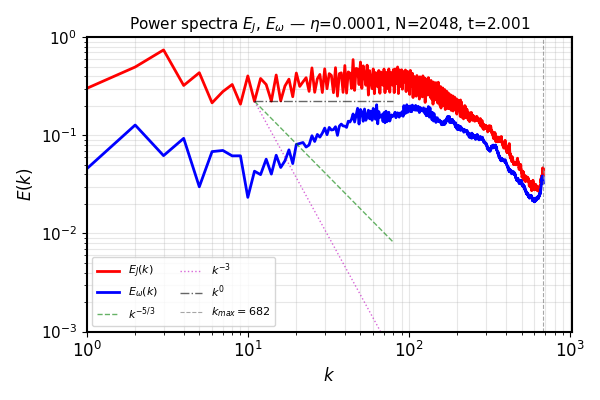}
\caption{Same as Figure~\ref{fig:plasmo_kmax32} but for $k_{\rm sup} = 128$.
The larger spectral content of the perturbation results in a significantly
higher number of plasmoids ($N_p = 22$) compared to the $k_{\rm sup} = 32$
case ($N_p = 9$), while the power spectrum at $t = 2.0$ confirms that the
simulation remains well-resolved.
$N = 2048$, $\eta = 10^{-4}$, $t^* = 1.5$, $\varepsilon = 10^{-3}$.}
\label{fig:plasmo_kmax128}
\end{figure}

The development of plasmoids for $k_{\rm sup} = 32$ and $k_{\rm sup} = 128$ is
illustrated in Figures~\ref{fig:plasmo_kmax32} and \ref{fig:plasmo_kmax128}
respectively, both for $t^* = 1.5$ and $\varepsilon = 10^{-3}$. The contrast
between the two cases is striking: $k_{\rm sup} = 128$ produces a much denser
plasmoid chain, while the power spectra in both cases confirm that the simulations
remain well-resolved throughout, with no spurious thermalisation at large $k$.

\begin{figure*}
\centering
\includegraphics[width=0.48\textwidth]{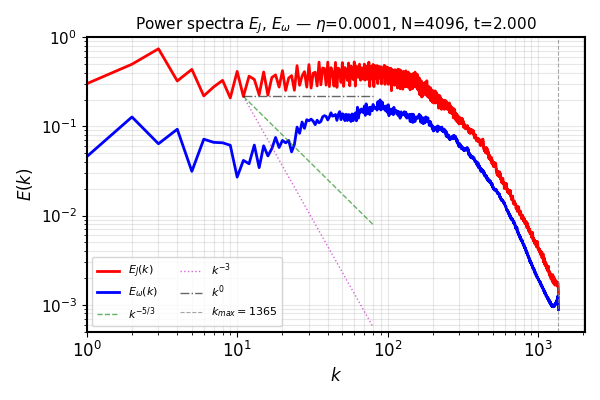}
\includegraphics[width=0.48\textwidth]{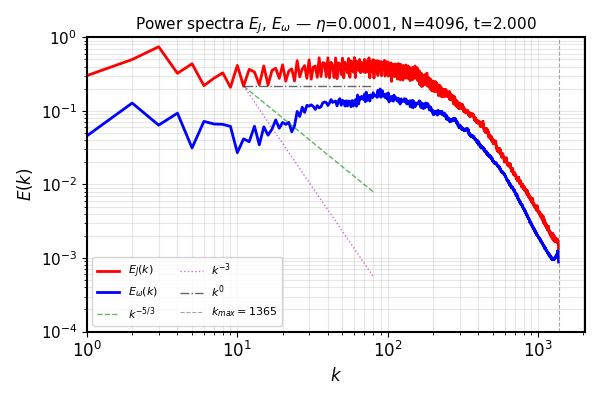}
\caption{Power spectra $E_J(k)$ at $N = 4096$ for $k_{\rm sup} = 32$ (left) and
$k_{\rm sup} = 128$ (right), at $t^* = 1.5$, $\varepsilon = 10^{-3}$,
$\eta = 10^{-4}$. Both spectra satisfy the convergence criterion and display a
well-developed dissipative cascade, confirming that the physical plasmoids are
fully resolved and that the results obtained at $N = 2048$ are robust. The
corresponding current sheet structures are not shown as they are virtually
indistinguishable from the $N = 2048$ cases (Figures~\ref{fig:plasmo_kmax32}
and \ref{fig:plasmo_kmax128}).}
\label{fig:conv4096}
\end{figure*}

To further confirm the robustness of these results, we have repeated the two
representative cases ($k_{\rm sup} = 32$ and $k_{\rm sup} = 128$, $t^* = 1.5$,
$\varepsilon = 10^{-3}$) at doubled resolution $N = 4096$.
The plasmoid counts differ by at most $1$ for $k_{\rm sup} = 32$ and by at most $2$
for $k_{\rm sup} = 128$ with respect to the $N = 2048$ results, well within the
$10\%$ counting uncertainty, confirming that the number of plasmoids is fully
converged. The corresponding current sheet structures are nearly
indistinguishable from the $N = 2048$ cases and are therefore not shown.
More importantly, Figure~\ref{fig:conv4096} shows the power spectra
$E_J(k)$ and $E_\omega(k)$ at $N = 4096$ for both cases side by side. Both spectra satisfy the
convergence criterion with a large margin, decreasing by more than two decades before
reaching $k_{\max}$. This confirms that the physical plasmoids reported here are
not resolution artifacts, and that the turbulent cascade induced by the plasmoid
activity is fully resolved at $N = 2048$.

\section{Comparison with theory and previous simulations}
\label{sec:comparison}

We now compare our results with the theoretical predictions of \citet{Comisso2017}
and the numerical simulations of \citet{Huang2017}, both of which provide
quantitative estimates of the growth rate $\gamma \tau_A$ and number of plasmoids
$N_p$ (via the dominant wavenumber $k_d$) for forming current sheets at Lundquist
numbers comparable to ours. Before comparing with previous works, we note that
\citet{Comisso2017} and \citet{Huang2017} normalize times to $\tau_A = L/v_A$
where $L = L_{cs}/2$ is the current sheet half-length, whereas we use the full
length $L_{cs}$. All values reported in Table~\ref{tab:comparison} have been
converted to our normalization, implying a factor of 2 with respect to the
original references.

\citet{Comisso2017} developed a general theory of the plasmoid instability in
forming current sheets based on a principle of least time, yielding predictions
for the dominant wavenumber and growth rate as functions of $S$ and the noise
amplitude $\varepsilon$. For $S \sim 10^5$ and representative noise levels, their
Figures 6 and 7 give a dominant growth rate $\gamma \tau_A \simeq 10$--$20$
and a dominant wavenumber $k_d L_{cs} \simeq 100$--$200$ in our normalization.
These values are in good agreement with our best-resolved results
($\gamma \tau_A \simeq 12$, $k_d L_{cs} \simeq 180$, $N_p \simeq 28$),
confirming the consistency of our simulations with the theoretical predictions.
These theoretical predictions have been confronted with direct numerical
simulations by \citet{Huang2017}, using the coalescence instability
configuration. At their case S1 ($S = 1.25 \times 10^5$), they measured a
disruption growth rate $\gamma \tau_A \simeq 6$ and a dominant wavenumber
$k_d L_{cs} \simeq 50$ in our normalization. \rev{The less favorable agreement
with the theory of \citet{Comisso2017} in their case can be attributed to the
presence of a sustained reconnection outflow in their coalescence setup, which
stretches the wavelengths of unstable modes and introduces an advective loss,
reducing both $\gamma_d$ and $k_d$ \citep{Huang2019}. Our Orszag-Tang
configuration, by contrast, produces a transient current sheet in which
some reconnection outflow develops during the thinning phase but ceases when the
sheet retracts after the peak of current density. This limits the time available
for the outflow to affect the mode selection, which explains the better agreement
with the theory of \citet{Comisso2017} that does not account for outflow effects.}

\rev{We emphasize that the dominant wavenumber $k_d$ reported in
Table~\ref{tab:comparison} is obtained from the number of plasmoids counted
along the current sheet direction ($x$), and thus corresponds to $k_x$, which
is directly comparable to the theories of \citet{Comisso2017} and
\citet{Huang2017}. It should not be confused with the peak of the radial power
spectrum at $k \simeq 80$--100, which receives contributions from both the
spacing of structures along the sheet ($k_x$) and their transverse confinement
($k_y \sim 1/\delta$), whose relative importance depends on the specific
parameters.}

Table~\ref{tab:comparison} summarizes this comparison.
\begin{table*}
\centering
\caption{Comparison of growth rate $\gamma \tau_A$, dominant wavenumber $k_d L_{cs}$
and number of plasmoids $N_p$ between theoretical predictions, previous simulations,
and the present work, at $S \sim 10^5$.
\label{tab:comparison}}
\begin{tabular}{lccccc}
\hline\hline
Study & $S$ & $\gamma \tau_A$ & $k_d L_{cs}$ & $N_p$ & Method \\
\hline
\citet{Comisso2017} (theory)             & $\sim 10^5$      & $10$--$20$                & $100$--$200$ & ---  & least-time principle \\
\citet{Huang2017}                   & $1.25\times10^5$ & $\simeq 6$       & $\simeq 50$  & ---  & coalescence, DNS     \\
This work ($t^*=1.9$, $k_{\rm sup}=256$) & $\sim 10^5$      & $\simeq 12$ & $\simeq 180$ & $28$ & Orszag-Tang, DNS     \\
\hline
\multicolumn{6}{l}{\small For the present work, $\tau_A = L_{cs}/v_A \simeq 1$ since $L_{cs} \simeq 3$ and $v_A \simeq 3$.} \\
\multicolumn{6}{l}{\small All values taken from Comisso et al. and Huang et al. converted to our normalization (see text).} \\
\end{tabular}
\end{table*}
The overall agreement supports the interpretation that the effective growth rate
and plasmoid number measured in our simulations are consistent with the
\citet{Comisso2017} theory for forming current sheets.

\subsection{Continuous perturbation}
\label{sec:continuous}

The perturbation procedure employed in the previous sections, where noise is
injected at a single instant $t^*$, is admittedly somewhat artificial. It was
adopted because it greatly facilitates the diagnostics, in particular the
measurement of growth rates and the identification of the role of each parameter
($t^*$, $\varepsilon$, $k_{\rm sup}$), and because it provides a clear
explanation of the GM\&A results where the perturbation applied at $t = 0$ is
too early to trigger the instability.

A more physically motivated approach consists in applying a continuous random
perturbation at every time step, mimicking the persistent background noise that
is always present in natural systems and in non-spectral numerical codes. To
this end, we have performed additional simulations in which a random noise of
amplitude $\varepsilon_{\rm cont}$, filtered to wavenumbers $k \leq k_{\rm sup}$,
is added to the vector potential $A$ at each time step.

The results are qualitatively similar to those obtained with the single-injection
procedure at $\varepsilon = 10^{-3}$, but the critical amplitude required to
trigger physical plasmoids is significantly lower: continuous perturbation
amplitudes of $\varepsilon_{\rm cont} \sim 10^{-5}$--$10^{-4}$ are sufficient
to produce similar well-developed plasmoid chains, i.e.\ one to two orders of magnitude
below the single-injection threshold. The measured growth rates are essentially
unchanged, while the number of plasmoids is reduced by approximately $20\%$
compared to the single-injection case at the same $k_{\rm sup}$. This modest
reduction is consistent with the fact that the single-injection procedure
artificially forces all unstable modes simultaneously at $t^*$, whereas
continuous noise seeds the modes more gradually, allowing the dominant mode to
emerge more naturally from the competition.

These results confirm that the triggering conditions identified in the previous
sections, namely the requirement for sufficient noise amplitude and appropriate
spectral content, are robust and not an artifact of the single-injection
procedure.

\section{Conclusions}
\label{sec:conclusion}

We have performed a systematic numerical study of the conditions for triggering
the plasmoid instability in a forming current sheet, using a pseudo-spectral code
applied to the Orszag-Tang vortex configuration at $S \sim 10^5$. Our results
address two central questions raised in the introduction: how to distinguish
physical from spurious plasmoids (driven by insufficient resolution), and under
what conditions physical plasmoids can be triggered.


\textit{Physical versus spurious plasmoids.} A first result of this study is the
confirmation and extension of the spectral diagnostic introduced by
GM\&A. The power spectrum criterion $\max_k\{E_J(k)\} \geq
10\,E_J(k_{\max})$ reliably identifies well-resolved simulations, and we have
shown that it remains valid even in the presence of physical plasmoids, where the
spectrum is significantly more structured than in the quiescent case. The
complementary use of the vorticity spectrum $E_\omega(k)$, which exhibits the
same under-resolution signature simultaneously, provides an independent
confirmation and strengthens the diagnostic. The measured growth rates and plasmoid
numbers are in good agreement with the theoretical predictions of \citet{Comisso2017}
and the simulations of \citet{Huang2017}, consistently with the physical nature of
the plasmoids obtained under the right triggering conditions. We therefore recommend
the systematic use of both $E_J(k)$ and $E_\omega(k)$ in future reconnection
studies, in particular when plasmoids are observed.


\textit{Triggering conditions and link with code type.} A second central result,
obtained in the specific context of the Orszag-Tang vortex at $S \sim 10^5$, is
that physical plasmoids can be triggered in a well-resolved spectral simulation,
but only when three conditions are simultaneously met: the perturbation must be
applied sufficiently late (close to the peak of the current density), with an
amplitude above a critical threshold $\varepsilon_c \sim 10^{-5}$, and with a
spectral content $k_{\rm sup}$ exceeding the dominant unstable wavenumber. We note that
this finite-time constraint is intimately linked to the fact that the current sheet
forms on an Alfv\'enic timescale ($t_{\rm peak} \simeq 1.9 \sim \tau_A$), which
is precisely the regime considered by \citet{Comisso2017}. This explains
the apparent paradox raised by GM\&A: in their simulations, the
perturbation was applied at $t = 0$, too early for the instability to develop
within the finite lifetime of the current sheet. \rev{More broadly, this result
highlights the importance of careful resolution studies in finite-difference and
finite-element codes, where the intrinsic truncation noise is always present and
could exceed the critical threshold $\varepsilon_c$. In such codes, plasmoids
may form spontaneously, but this does not guarantee their physical nature:
convergence with respect to grid size should be verified to ensure that the
observed structures are not favored by insufficient resolution. The spectral
diagnostics $E_J(k)$ and $E_\omega(k)$ proposed in this work could provide a
useful tool for this purpose.} In a spectral code, on the contrary, an explicit
and well-timed perturbation is necessary, at least in our setup. We have further
verified that a continuous random perturbation applied at every time step yields
similar results, with a critical amplitude one to two orders of magnitude lower
than in the single-injection case, confirming the robustness of the identified
triggering conditions.


\rev{\textit{Physical origin of the noise.} An important question raised by the
present study is: what physical process provides the required noise in a real
system? Several candidates can be invoked. In astrophysical and laboratory
plasmas, kinetic effects (particle noise, micro-instabilities,
etc.) become relevant when the current sheet thins down to scales
approaching kinetic lengths (electron and ion skin depths, ion Larmor radius, etc.) and would naturally provide a source of broadband
noise at the relevant time. In a turbulent environment, ambient fluctuations
continuously perturb the forming current sheet, providing a noise source analogous
to our continuous perturbation protocol. A definitive identification of the
dominant physical noise source goes beyond the scope of this MHD study, but the
present results provide a quantitative framework for assessing whether a given
noise level is sufficient to trigger the plasmoid instability within the finite
lifetime of a forming current sheet.}


\textit{Open questions.} Two important questions are left open by the present
study. First, whether the distinction between spurious and physical plasmoids has
a significant impact on the measured reconnection rate. If spurious plasmoids
artificially trigger the fast reconnection regime, the measured rate could be
significantly affected. Addressing this would require simulations over a range of
$S$ values in the plasmoid regime, which is beyond the scope of the present study.
Second, the present results were obtained in the specific context of the
Orszag-Tang vortex, where the current sheet forms and relaxes on a dynamical
timescale. An important extension would be to other setups representative of
quasi-stationary reconnection configurations, where the current sheet persists on
a much longer timescale, relaxing the finite-time constraint on plasmoid
triggering. Both directions represent important avenues for future work.

\section*{Acknowledgments}
The author thanks Alexandros Alexakis for valuable discussions and constructive
feedback that have helped improve this work. \rev{The author also thanks the
anonymous referees for their insightful comments and suggestions.}

\appendix
\section{Measurement procedures for $\gamma$ and $N_p$}
\label{app:diagnostics}

The growth rate $\gamma$ reported throughout this paper is measured by monitoring
the amplitude of the magnetic vector potential perturbation, defined as
\begin{equation}
\delta A(t) = \max_{\mathbf{x}} |A(\mathbf{x},t) - A_0(\mathbf{x},t)|,
\end{equation}
where $A_0(\mathbf{x},t)$ is the unperturbed (reference) solution at the same
time. During the linear growth phase, $\delta A(t)$ grows exponentially as
$\delta A \propto e^{\gamma t}$, and $\gamma$ is extracted from the slope of
$\ln \delta A(t)$ over the interval during which this growth is well-defined.

It must be emphasized that $\gamma$ as measured here is an effective growth rate,
not the classical linear growth rate of a static Sweet-Parker current sheet. For
a forming current sheet, the background equilibrium is itself time-dependent: the
sheet thins, reaches a peak, and then relaxes on a timescale comparable to the
instability growth time. As a result, there is no well-defined normal mode, and
$\gamma$ reflects the integrated amplification of the perturbation over the finite
lifetime of the unstable phase. This effective rate is the relevant quantity for
comparison with the theoretical predictions of \citet{Comisso2017} and the
numerical results of \citet{Huang2017}, as discussed in
Section~\ref{sec:comparison}.

\begin{figure*}
\centering
\includegraphics[width=0.99\textwidth]{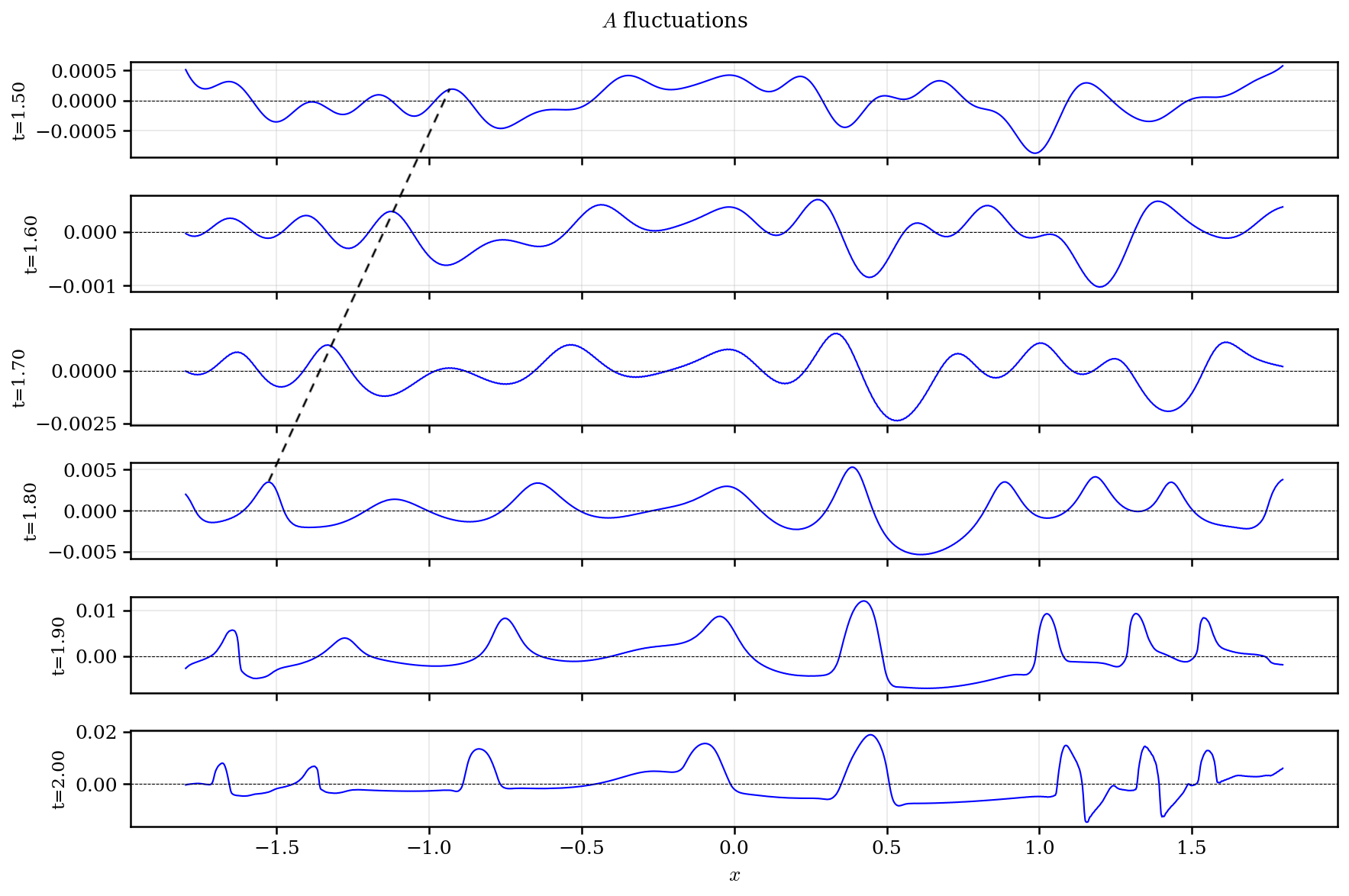}
\caption{\rev{Fluctuations of the magnetic vector potential $A(x, y=0)$ around a
low-order polynomial fit, at six successive times for the case $t^* = 1.5$,
$\varepsilon = 10^{-3}$, $k_{\rm sup} = 32$. \textcolor{red}{The number of distinct oscillations ($\sim 9$) remains approximately
stable during the growth phase ($t = 1.6$--$1.8$) with a modest decrease
due to advection toward the sheet endpoints}, confirming that the dominant wavenumber $k_d$ is
well-defined before disruption ($t \simeq 1.8$). The dashed line tracks a
single fluctuation across successive panels, illustrating its advection
toward the sheet endpoint. At later times ($t = 1.9$--$2.0$), the islands are advected toward the sheet
endpoints, reducing the number of structures. Times are in units of the
Alfv\'en time $\tau_A$ (see text). $N = 2048$, $\eta = 10^{-4}$.}}
\label{fig:Afluct}
\end{figure*}

\rev{The number of plasmoids $N_p$ (and hence the dominant wavenumber
$k_d = 2\pi N_p / L_{cs}$) is measured using two independent methods: (i)~visual
counting of current density enhancements just before disruption, and
(ii)~counting the oscillations of the magnetic vector potential fluctuations
$A(x, y=0)$ around a low-order polynomial fit at several times during the growth
phase. The two methods agree to within 10\%. Figure~\ref{fig:Afluct} shows these
fluctuations at six successive times for a representative case ($t^* = 1.5$,
$\varepsilon = 10^{-3}$, $k_{\rm sup} = 32$). \textcolor{red}{The number of oscillations ($\sim 9$) remains approximately stable
during the growth phase up to disruption ($t \simeq 1.8$) with a modest
decrease due to advection toward the sheet endpoints. The spacing between
peaks, and hence $k_d$, remains well-defined before the nonlinear regime
sets in.} At later times ($t = 1.9$--$2.0$), the islands are advected toward the sheet
endpoints, reducing the number of structures.}

\bibliographystyle{aasjournal}

\begin{thebibliography}{}

\bibitem[Garc\'ia Morillo \& Alexakis(2025)]{Alexakis2025}
Garc\'ia Morillo, J.M., \& Alexakis, A.\ 2025, J. Fluid Mech., 1007, R3

\bibitem[Biskamp(2000)]{Biskamp2000}
Biskamp, D.\ 2000, Magnetic Reconnection in Plasmas, Cambridge University Press

\bibitem[Bhattacharjee et al.(2009)]{Bhattacharjee2009}
Bhattacharjee, A., Huang, Y.-M., Yang, H., \& Rogers, B.\ 2009, Phys. Plasmas, 16, 112102

\bibitem[Comisso et al.(2016)]{Comisso2016}
Comisso, L., Lingam, M., Huang, Y.-M., \& Bhattacharjee, A.\ 2016, Phys. Plasmas, 23, 100702

\bibitem[Comisso \& Grasso(2016)]{ComissoGrasso2016}
Comisso, L., \& Grasso, D.\ 2016, Phys. Plasmas, 23, 032111

\bibitem[Comisso et al.(2017)]{Comisso2017}
Comisso, L., Lingam, M., Huang, Y.-M., \& Bhattacharjee, A.\ 2017, ApJ, 850, 142

\bibitem[Huang \& Bhattacharjee(2010)]{HuangBhattacharjee2010}
Huang, Y.-M., \& Bhattacharjee, A.\ 2010, Phys. Plasmas, 17, 062104

\bibitem[Huang et al.(2017)]{Huang2017}
Huang, Y.-M., Comisso, L., \& Bhattacharjee, A.\ 2017, ApJ, 849, 75

\bibitem[Huang et al.(2019)]{Huang2019}
Huang, Y.-M., Comisso, L., \& Bhattacharjee, A.\ 2019, Phys. Plasmas, 26, 092112

\bibitem[Loureiro et al.(2007)]{Loureiro2007}
Loureiro, N.F., Schekochihin, A.A., \& Cowley, S.C.\ 2007, Phys. Plasmas, 14, 100703

\bibitem[Loureiro \& Uzdensky(2016)]{LoureiroPPCF2016}
Loureiro, N.F., \& Uzdensky, D.A.\ 2016, Plasma Phys. Control. Fusion, 58, 014021

\bibitem[Ng et al.(2008)]{Ng2008}
Ng, C.S., Rosenberg, D., Germaschewski, K., Pouquet, A., \& Bhattacharjee, A.\ 2008, ApJS, 177, 613

\bibitem[Orszag \& Tang(1979)]{Orszag1979}
Orszag, S.A., \& Tang, C.-M.\ 1979, J. Fluid Mech., 90, 129

\bibitem[Priest \& Forbes(2000)]{Priest2000}
Priest, E.R., \& Forbes, T.G.\ 2000, Magnetic Reconnection: MHD Theory and Applications, Cambridge University Press

\bibitem[Pucci \& Velli(2014)]{Pucci2014}
Pucci, F., \& Velli, M.\ 2014, ApJL, 780, L19

\bibitem[Samtaney et al.(2009)]{Samtaney2009}
Samtaney, R., Loureiro, N.F., Uzdensky, D.A., Schekochihin, A.A., \& Cowley, S.C.\ 2009, Phys. Rev. Lett., 103, 105004

\bibitem[Tenerani \& Velli(2020)]{Tenerani2020}
Tenerani, A., \& Velli, M.\ 2020, MNRAS, 491, 4267

\end{thebibliography}

\end{document}